%% file: main.tex
\documentclass[sigconf,anonymous=False,natbib=true]{acmart}

\AtBeginDocument{%
  }


\input{preamble/packages.tex}

\input{preamble/definitions.tex}

\begin{document}

\title{IRLab@iKAT24: Learned Sparse Retrieval with Multi-aspect LLM Query Generation for Conversational Search}

\input{preamble/authors.tex}

\input{sections/00_abstract}

\input{sections/keywords}

\maketitle

\renewcommand{\thefootnote}{\fnsymbol{footnote}}
\footnotetext[1]{Equal contribution.}
\renewcommand{\thefootnote}{\arabic{footnote}}

\input{sections/01_introduction}

\input{sections/02_tasks}

\input{sections/03_methodology}

\input{sections/04_experimental_setup}

\input{sections/05_results}
\input{sections/06_conclusions}

\bibliographystyle{ACM-Reference-Format}
\balance
\bibliography{main}

\input{sections/07_appendix}

\end{document}

%% file: preamble/packages.tex
\usepackage[inline]{enumitem}
\usepackage{pbox}
\usepackage{balance}
\usepackage{todonotes}
\newlist{inlinelist}{enumerate*}{1}
\setlist*[inlinelist,1]{label=\roman*),itemjoin={{, }},itemjoin*={{, and }}}
\usepackage{url}
\usepackage[normalem]{ulem}
\usepackage{multirow}
\usepackage{multicol}
\usepackage{booktabs}
\usepackage{tikz}
\usepackage{xcolor}
\usepackage{lipsum}
\usepackage{acronym}
\usepackage{subcaption}
\usepackage[inline]{enumitem}
\usepackage{xspace}
\usepackage{arydshln}
\usepackage{array}
\usepackage{graphicx}
\usepackage{tabularx}

\usepackage{newfloat}
\DeclareFloatingEnvironment[fileext=lop, listname={List of prompts}, 
                            name=Prompt, placement=h]{prompt}

\usepackage{listings}
\usepackage{xcolor}

%% file: preamble/definitions.tex
\acrodef{CS}{Conversational Search}
\acrodef{CSA}{Conversational Search Agent}
\acrodef{PTKB}{Personal Text Knowledge Base}
\acrodef{TREC}{TExt Retrieval Conference}
\acrodef{iKAT}{Interactive Knowledge Assistance Track}
\acrodef{CAsT}{Conversational Assistance Track}
\acrodef{NIST}{National Institute of Standards and Technology}
\acrodef{LLM}{Large Language Model}
\acrodef{LSR}{Learned Sparse Retrieval}

\acrodef{IR}{Information Retrieval}
\acrodef{NLP}{Natural Language Processing}
\acrodef{PEFT}{Parameter-Efficient Fine-Tuning}
\acrodef{ICL}{In-Context Learning}
\acrodef{LoRA}{Low-Rank Adaptation}

\acrodef{CQR}{Conversational Query Rewriting}

\acrodef{MSE}{Mean Square Error}

\newcolumntype{P}[1]{>{\centering\arraybackslash}p{#1}}

%% file: preamble/authors.tex
\author{Simon Lupart\footnotemark[1]}
\affiliation{%
  \institution{University of Amsterdam}
  \city{Amsterdam}
  \country{Netherlands}}
\email{s.c.lupart@uva.nl}

\author{Zahra Abbasiantaeb\footnotemark[1]}
\affiliation{%
  \institution{University of Amsterdam}
  \city{Amsterdam}
  \country{Netherlands}}
\email{z.abbasiantaeb@uva.nl}

\author{Mohammad Aliannejadi}
\affiliation{%
  \institution{University of Amsterdam}
  \city{Amsterdam}
  \country{Netherlands}}
\email{m.aliannejadi@uva.nl}

\renewcommand{\shortauthors}{S. Lupart, Z. Abbasiantaeb and M. Aliannejadi}

%% file: sections/00_abstract.tex
\begin{abstract}

The Interactive Knowledge Assistant Track (iKAT) 2024 focuses on advancing conversational assistants, able to adapt their interaction and responses from personalized user knowledge. The track incorporates a Personal Textual Knowledge Base (PTKB) alongside Conversational AI tasks, such as passage ranking and response generation. Query Rewrite being an effective approach for resolving conversational context, we explore Large Language Models (LLMs), as query rewriters. Specifically, our submitted runs explore multi-aspect query generation using the MQ4CS framework, which we further enhance with Learned Sparse Retrieval via the SPLADE architecture, coupled with robust cross-encoder models. We also propose an alternative to the previous interleaving strategy, aggregating multiple aspects during the reranking phase. Our findings indicate that multi-aspect query generation is effective in enhancing performance when integrated with advanced retrieval and reranking models. Our results also lead the way for better personalization in Conversational Search, relying on LLMs to integrate personalization within query rewrite, and outperforming human rewrite performance.


\end{abstract}

%% file: sections/keywords.tex


\keywords{conversational search, personalization, retrieval augmented generation, query understanding, neural sparse retrieval}

%% file: sections/01_introduction.tex
\section{Introduction}

With the rapid advancement of conversational AI systems, conversational search (CS) has seen an important development recently, along with a need for systems that can engage in meaningful, context-aware interactions. The TREC interactive Knowledge Assistant Track (iKAT) 2024 specifically addresses these needs for a second year now~\cite{ikat23}, by developing conversational assistants capable of discussing a wide range of topics with users. These assistants are not only designed to engage in dialogue but also to support users in making informed decisions by providing contextually relevant, personalized responses and dynamically ranked information within each conversational exchange.

The TREC iKAT achieves this by leveraging two main sources of knowledge: a broad knowledge collection and a Personal Text Knowledge Base (PTKB), which contains user-specific information. This PTKB captures individual preferences, constraints and prior experience of the user, all in a textual format. TREC iKAT builds upon the TREC Conversational Assistance Track (CAsT)~\cite{cast19,cast20,cast22}, extending its core task to incorporate personalization. By integrating user-centric data, the track makes the retrieval and response generation processes significantly more challenging, as systems must adapt to each user's unique context and preferences.
To address this complexity, TREC iKAT includes several subtasks: passage ranking, PTKB classification, and response generation. To help understand the behavior of the different models, the track also proposes different submission formats: (1) an automatic run, where all tasks are processed together by the system, (2) a manual run, which includes human-generated rewrites 
of each conversational turn to enhance context modeling for retrieval, and (3) a generation-only run, which concentrates on response generation using an already ranked list of passages provided by the organizers. 

Recent studies in conversational search have demonstrated that Large Language Models (LLMs) are effective in tackling such tasks~\cite{mq4cs,mo2024chiqcontextualhistoryenhancement,mao2023largelanguagemodelsknowllm4cs,ikat23,lupart2024disco}, disambiguating the user's latest utterance in the conversation. Building on this, our submitted runs explore advanced query rewriting techniques to resolve conversational ambiguities. Our work leverages the recent MQ4CS framework~\cite{mq4cs}, which we further extend. Additionally, we aim to push the state of the art in conversational search by deploying enhanced retrieval and reranking models, strengthening the accuracy and relevance of system-generated responses across personalized conversations.

Through our runs, we show how the MQ4CS framework generalizes to the new iKAT 2024 benchmark. We also propose a variation of the original MQ4CS framework and show the additional improvement of combining it with state-of-the-art retrieval and reranking models.

%% file: sections/02_tasks.tex
\section{Tasks}

As last year, iKAT 2024 provides user utterance, conversation history, and Personal Text Knowledge Base (PTKB) as input for each conversational turn. The conversation history includes the user's previous user utterances and gold responses from the system. TREC iKAT 2024 benchmark includes 103 turns across 13 topics, each paired with a unique user persona (i.e., PTKB) of ~16.8 descriptive statements on average. Passage ranking is done on the ClueWeb-iKAT collection, with each conversational turn evaluated independently. The iKAT subtasks are as follows:
\begin{enumerate}[leftmargin=*]
    \item Passage Ranking: aims to retrieve and rank the relevant passages from the given collection in response to each user utterance. To understand the user query per each turn, the system must do reasoning over relevant statements from PTKB and conversation history.
    \item Response Generation: The goal of this task is to provide a natural answer to the user utterance, following the flow of the conversation, and based on the retrieved passages. The responses must be grounded from the retrieved passages from the collection. Hence, the responses must be generated using at least one passage, referred to as “provenance”, from the source collection.
    \item PTKB Classification: the PTKB consists of several statements. The goal of this task is to classify the statements within the PTKB as relevant/irrelevant for each conversational turn. The output of this task is a binary label for each statement from the PTKB.
\end{enumerate}

Besides, the organizers proposed different types of runs: first the automatic runs (including the three subtasks all at once), then the manual runs (where user utterances are disambiguated from the conversation history and already include relevant PTKB statements), and finally the response generation-only subtask. We focus on the automatic runs, as the most challenging run, and also on the manual runs.

%% file: sections/03_methodology.tex
\section{Methodology}
In this section we explain the method used for the different tasks including (1) Passage ranking, (2) Response generation and (3) PTKB classification.

\subsection{Passage Ranking}
Our proposed models for passage ranking is based on the MQ4CS~\cite{mq4cs} framework. We further make some changes to the rank fusion step of MQ4CS framework in our submissions. We will briefly explain the MQ4CS framework and our modification to the rank fusion step in the following. 




\textbf{Multiple Aspect Query Generation.}
We strongly rely on LLMs to disambiguate the conversational context and persona of the user. In particular, we use the MQ4CS framework proposed by ~\cite{mq4cs}, generating multiple queries each covering different aspects of the information need. 
MQ4CS proposes breaking the user information need into multiple queries rather than forming a single query rewrite which includes the complete information need.
This allows for better coverage of the collections, as a query expansion, but also enables a decomposition of the query into several sub-pieces of simpler information needs. 
Note that the queries would be generated for each user utterance by doing reasoning over both the context and personal information about the user provided in PTKB.

\begin{figure}[t]
    \centering
    \includegraphics[width=0.95\linewidth]{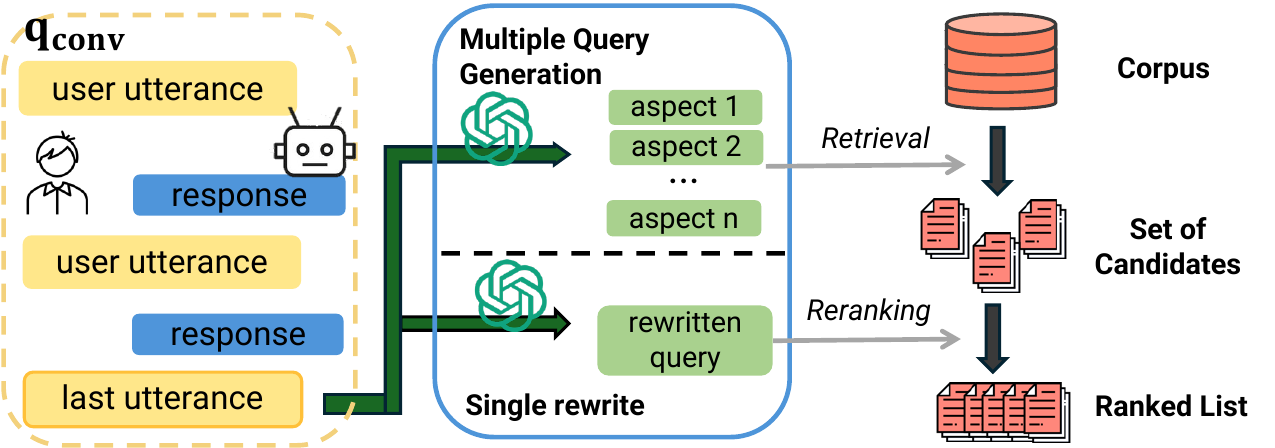}
    \caption{MQ4CS-QR, as a variation of the original MQ4CS framework, that uses multiple query generation for \textit{Retrieval}, but an independent single rewrite for \textit{Reranking}.}
    \label{fig:diag}
\end{figure}

\textbf{Rank Fusion.} 
In MQ4CS, a retrieval and reranking pipeline is applied to each of the generated queries, producing several ranked lists that will be interleaved into a final ranking.
As mentioned by the authors, this interleaving is sub-optimal. Hence, we propose to only use the multiple generated queries for retrieval and then re-rank all retrieved passages using a single query rewrite. To be more specific, instead of interleaving the passages retrieved by multiple generated queries, we re-rank them using a single query rewrite as described in Figure~\ref{fig:diag}. We follow the zero-shot query-rewriting baseline based on the GPT-4 model proposed by \cite{mq4cs} to generate a single query rewrite. This would benefit from a large recall, as the different queries cover different aspects, and using these queries ensures having a diverse set of passages from different sources of information, but still have high precision, as relying on a single query rewrite to re-rank them.


\textbf{Retrieval \& Re-ranking.} We extend the original work on MQ4CS that uses term-based retrieval with learned sparse retrieval. Our runs use SPLADE, which adapts the MLM pre-trained layer of BERT to predict term importance and token expansion within the WordPiece vocabulary~\cite{spladev1,formal2021splade,10.1145/3477495.3531857/splade++,lassance2024spladev3}.

For reranking, we also propose to use more effective re-rankers compared to the one used in the original MQ4CS paper. Those re-rankers were trained with negatives from SPLADE, making them more effective in detecting noisy passages that could have been retrieved from our first-stage model. We also rely on the ensembling of rankers, using a min-max normalization on the produced ranked list.




\subsection{Response Generation}

For response generation, we rely on the simple yet effective Retrieval Augmented Generation (RAG) pipeline using LLMs on top of retrieved passages~\cite{RAG-first-work}. In this pipeline, the model is given the user utterance, context of the conversation, PTKB of the user, and the top passages retrieved from our passage ranking. Each submission follows this pipeline, with a different top ranking. We use a similar prompt to \citeauthor{abbasiantaeb2023llm}, including all PTKB within the prompt. This approach was effective for the last year of iKAT.

\input{tables/main_res}

\subsection{PTKB classification}

In both previous tasks (Ranking and Response Generation), all PTKB are included in the model's input. We thus entirely rely on LLMs to filter and extract relevant information from the user in different contexts. However, to better understand and explicit the model reasoning, we also prompted the model to classify relevant statements from the PTKB list. In particular, we follow the zero-shot approach proposed by \cite{abbasiantaeb2023llm} for iKAT 2023, where an LLM is prompted to return the relevant PTKB given the user utterance and context of the conversation, with an adaptation for classification.,  

As an alternative approach, one could only use those filtered PTKB for the context modeling and response generation tasks, breaking down the personalization task, but we believe that LLMs are effective enough to combine multiple tasks.


%% file: tables/main_res.tex
\begin{table*}[t]
\centering
\caption{Experimental results on the official evaluation set. Bold and underlined indicate the best and second-best results across both Automatic and Manual runs.}
\label{tab:main_tab}
\begin{tabular}{l|l|l|l|cccccc}
\midrule
\textbf{Runs} & \textbf{LLM} & \textbf{Retrieval} & \textbf{Reranker} & \textbf{nDCG@5} & \textbf{nDCG} & \textbf{MRR} & \textbf{Recall@100} & \textbf{P@20} & \textbf{mAP} \\ 
\midrule
(1) MQ4CS-QR-deberta & GPT-4 & SPLADE & DebertaV3 & \underline{0.5213} & \underline{0.6098} & 0.8479 & 0.4366 & \underline{0.5927} & 0.3486 \\ 
(2) MQ4CS-QR-ensemble & GPT-4 & SPLADE & Ensemble & 0.5155 & \textbf{0.6192} & \textbf{0.8644} & \underline{0.4657} & \textbf{0.6184} & \textbf{0.3698} \\ 
(3) GPT4QR-deberta & GPT-4 & SPLADE & DebertaV3 & 0.5071 & 0.5420 & 0.8315 & 0.4135 & 0.5684 & 0.3166 \\ 
(4) GPT4QR-bm25-QD1 & GPT-4 & BM25 & MiniLM & 0.4768 & 0.4373 & 0.8476 & 0.3421 & 0.5199 & 0.2315 \\ 
\midrule
(5) HumanQR-deberta &Human& SPLADE & DebertaV3 & 0.4767 & 0.5496 & 0.7808 & 0.4263 & 0.5383 & 0.3072 \\
(6) HumanQR-ensemble &Human& SPLADE & Ensemble & \textbf{0.5384} & 0.5819 & \underline{0.8483} & \textbf{0.4777} & 0.5922 & \underline{0.3533} \\ 
\midrule
\end{tabular}
\end{table*}

%% file: sections/04_experimental_setup.tex
\section{Experimental Setup}

\subsection{Submitted Runs}
The final submitted runs can be decomposed into automatic and manual runs. We did not experiment further with the answer generation-only subtask.

\textbf{Automatic runs:}
\begin{enumerate}
\item \textit{MQ4CS-QR-deberta}: This run uses the MQ4CS framework, generating multiple queries ($\phi$=5, GPT4) for our first-stage SPLADE retrieval. For reranking, we generate an additional single query rewrite from GPT-4 and rerank it with a DebertaV3 cross-encoder. Both PTKB classification and response generation models use GPT-4 as LLM. All prompts are zero-shot, following MQ4CS~\cite{mq4cs}.



\item \textit{MQ4CS-QR-ensemble}: Similar to the first run, this submission relies on MQ4CS. The entire pipeline is similar except for re-ranking which uses an ensembling of 5 cross-encoder models (DebertaV2, DebertaV3, Roberta, Alert, Electra). Response generation and PTKB classification use the same method as in the previous run, with GPT-4.

\item \textit{GPT4QR-deberta}: For this run, we follow the common practice of using a single query rewrite. This rewrite is the same single rewrite used in both previous runs for reranking, generated from GPT-4 in zero-shot. Retrieval and reranking use SPLADE and DebertaV3.
Response generation and PTKB classification use the same method as in the previous runs, with GPT-4.

\item \textit{GPT4QR-bm25-QD1}: 
This run uses a single query, but with the prompting strategy of MQ4CS, with $\phi$=1, GPT-4. It further uses BM25 and a miniLM cross-encoder. Response generation and PTKB classification remain unchanged. The objective of this run was to compare it with the baseline runs from the organizers, to see the impact of the prompting strategies with a single query.


\end{enumerate}

\textbf{Manual Runs:}
\begin{enumerate}[resume]
\item \textit{HumanQR-deberta}: This manual run uses the single human rewrite for both retrieval and reranking. It relies on SPLADE and DebertaV3. Response generation and PTKB classification are added afterward from GPT-4.

\item \textit{HumanQR-ensemble}: similar to the first manual run, this submission uses the manual human rewrite, but with an ensemble of 5 cross-encoders for reranking. Response generation and PTKB classification are also added afterward from GPT-4.
\end{enumerate}

Note that all our submissions (both automatic and manual) use the same PTKB classification results.

\subsection{Hyperparameters}
We provide in this section the details and sources of the models we used. 
Our retrieval mostly relies on SPLADE, using the CoCondenser SelfDistil SPLADE++ checkpoint from HuggingFace\footnote{naver/splade-cocondenser-selfdistil}. We also rely on BM25 for one of the runs, with default hyperparameters.
For re-ranking, the list of rerankers is \textbf{debertav3 large}, \textbf{debertav2 xxlarge}~\cite{he2021debertav3,he2020deberta}, \textbf{roberta large}~\cite{liu2019roberta}, \textbf{albert v2-xxlarge}~\cite{lan2019albert} and \textbf{electra large}~\cite{clark2020electra}. All those models were fine-tuned by \citeauthor{lassance2023naver} for re-ranking. The list of rerankers can be found on HuggingFace\footnote{naver/trecdl22-crossencoder-\{debertav2, debertav3, electra, albert, roberta\}}. We also rely on the MS MARCO MiniLM cross-encoder\footnote{cross-encoder/ms-marco-MiniLM-L-6-v2}.
For ensembling, we follow the traditional setup with the ranx library. All reranking rerank the top 1000 of retrieved passages.

\subsection{Metrics}

We report traditional metrics in IR, that we received from TREC. This includes Mean Reciprocal Rank (MRR), Normalized Discounted Cumulative Gain (nDCG)~\cite{ndcg} and Precision, as well as recall and mean Average Precision. Those metrics are the ones used last year for iKAT~\cite{ikat23}. We however could not use a >=2 threshold for MRR and recall, as we did not have access to the qrels yet (only access to our results). For similar reason, we could also not evaluate the response generation, but we will update after we obtain the results from the organizers.








%% file: sections/05_results.tex
\section{Results}

In the section below, we present our retrieval results and main observations from the submitted runs.




\subsection{Retrieval Performance}

\textbf{Multiple Query Generation.} Table~\ref{tab:main_tab} contains the main retrieval results of our six submitted runs, manual and automatic. To first compare the performance when using multiple queries to represent the conversation, we look at the submitted runs (1) and (3), both reranked with DebertaV3. Results with the multiple queries show an important improvement in recall-oriented metrics, in particular a 2.3-point increase in Recall@100 and 3.2 points on mAP. This demonstrates the benefit of the MQ4CS framework combined with SPLADE. Also after the reranking, still comparing (1) and (3), we see that the improvement in recall for the first stage also leads to better precision after reranking. In particular, we see a 1.5-point gain on nDCG@5 and a 6.8-point increase on nDCG. This is possible because the set of candidate passages for re-ranking is of better quality.

\textbf{Human vs.\ Automatic.} Now looking at the manual runs, we see that the gap between manual and automatic runs is getting closer, with automatic runs outperforming the manual runs on nDCG, MRR, P@20, and mAP metrics. In particular when comparing (3) and (5) that both follow the same pipeline with a single rewrite either from Human or GPT4, we see how much GPT models now beat humans on the rewriting task. On the ensemble reranking approach ((2) and (6)), we also see that the automatic run outperforms on several metrics the manual one.

\textbf{Ensembling of Rerankers.}  Ensembling uses several cross-encoders and averages their ranked list, to improve robustness and effectiveness. Looking at the Human rewrite, we see that ensembling leads to better performance compared to using only DebertaV3. We observe a similar trend on automatic runs combined with MQ4CS ((1) and (2)), with an average 2 points increase on all metrics, except for the nDCG@5 where the ensembling is 0.005 points below the single re-ranker.

\subsection{Performance per Depths and Topics}

\begin{figure}[t]
    \centering
    \includegraphics[width=0.95\linewidth]{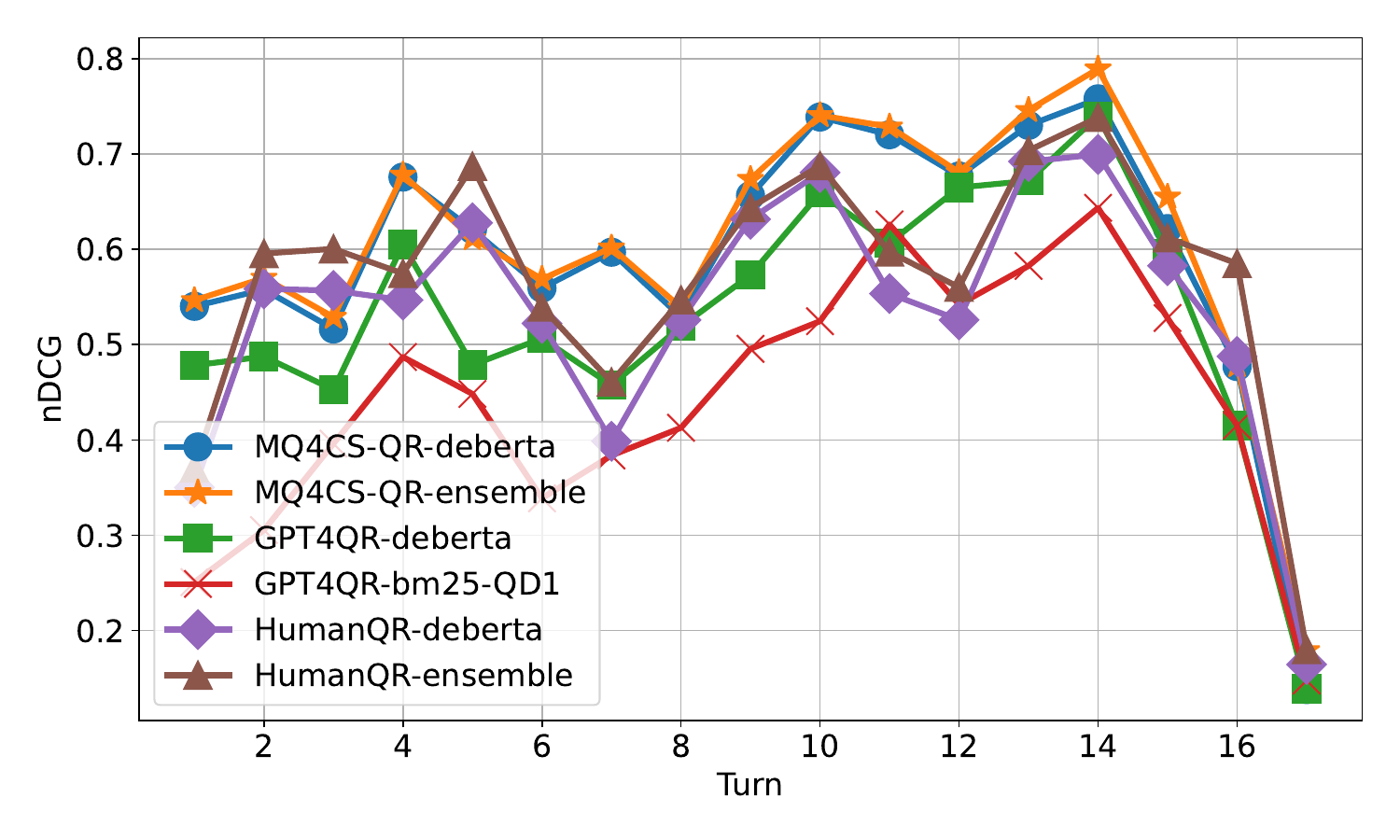}
    \caption{Performance per Depth.}
    \label{fig:depth}
\end{figure}

\textbf{Depth.} Figure \ref{fig:depth} presents the results of our set of models with respect to the depth of the turns of the conversation. While in previous years iKAT 2023 and generally in CAsT 2019-2022, longer conversations lead to harder context modeling, we see here that the nDCG increases as we go further within turns. An explanation for this can be the PTKB and the personalization task, as new information can be revealed by the user within the conversation, or relevant PTKB have been made explicit either by gold answers from previous turns or clarifying questions.

\begin{figure}[t]
    \centering
    \includegraphics[width=0.95\linewidth]{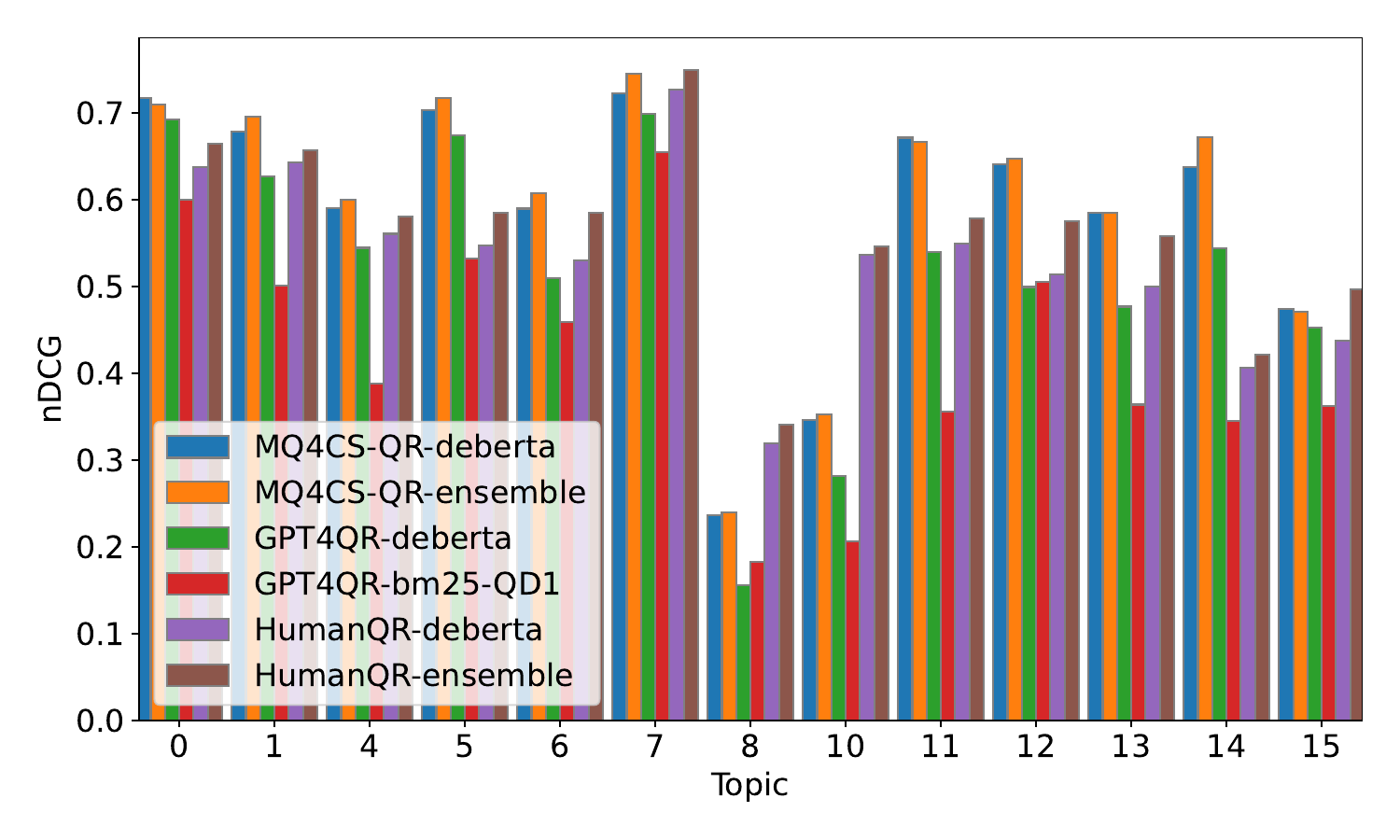}
    \caption{Performance per Topic.}
    \label{fig:topic}
\end{figure}

\textbf{Topics.} In Figure~\ref{fig:topic}, we plot the average nDCG on the different topics from iKAT. We see from this figure that some conversations are more challenging than others, in particular topic 8 on mountain bikes. Furthermore, we can identify topics 11 to 14, on which our best automatic runs MQ4CS-QR Deberta and ensemble (blue and orange bars) outperform the two manual runs (purple, brown).

%% file: sections/06_conclusions.tex
\section{Conclusion}

This report aims to describe the technical details of our iKAT 2024 submission. Through our submission, we demonstrate how 
the MQ4CS framework generalizes to iKAT 2024, and show the additional gains from combining it with Learned Sparse Retrieval and more effective cross-encoders models. We also propose a variation of the MQ4CS framework, that we will compare in future works to the original framework.


%% file: sections/07_appendix.tex
\newpage

\appendix

\section{Prompts}
\label{ap:prompt}
All our prompts re-use the one proposed within the MQ4CS~\cite{mq4cs} framework or previous work on iKAT 2023. We only changed the PTKB prompt, to adapt it to the classification task. Table~\ref{tbl:prompt-qd} recall the multiple query generation prompt, Table~\ref{tbl:prompt-rag} the RAG prompt for response generation, and Table~\ref{tbl:prompt-ptkb} the prompt for PTKB classification.


\begin{table}[h!]
\caption{The prompt designed for multiple QD using GPT-4 as a zero-shot learner.}
\centering
\begin{tabularx}{\linewidth}{X}
    \toprule
    \textbf{Multiple Query re-writing (QD).} \\
    \midrule
    \# Instruction: \textit{I will give you a conversation between a user and a system. Imagine you want to find the answer to the last user question by searching on Google. You should generate the search queries that you need to search on Google. Please don't generate more than \{$\phi$\} queries and write each query on one line.} \\
    \# Background knowledge: \{$ptkb$\} \\
    \# Context: \{$ctx$\} \\
    \# User question: \{$user\ utterance$\} \\
    \# Generated queries: \\
    \bottomrule
\end{tabularx}
\label{tbl:prompt-qd}
\end{table}

\begin{table}[h!]
\caption{The prompt designed for answer generation, using retrieved documents.}
\centering
\begin{tabularx}{\linewidth}{X}
    \toprule
    \textbf{Retrieval Augmented Answer Generation (RAG).} \\
    \midrule
    \# Doc1: \{$doc_{(1)}$\} \\
    \# Doc2: \{$doc_{(2)}$\} \\
    \# Doc3: \{$doc_{(3)}$\} \\
    \# Doc4: \{$doc_{(4)}$\} \\
    \# Doc5: \{$doc_{(5)}$\} \\
    \# \textit{I will give you a conversation between a user and a system. Also, I will give you some background information about the user. You should answer the last utterance of the user by providing a summary of the relevant parts of the given documents. Please remember that your answer shouldn't be more than 200 words.} \\
    \# Background information about the user: \{$ptkb$\} \\
    \# Conversation: \{$ctx$\} \\
    \# User query: \{$user\ utterance$\} \\
    \bottomrule
\end{tabularx}
\label{tbl:prompt-rag}
\end{table}

\begin{table}[h!]
\caption{The prompt designed for PTKB classification.}
\centering
\begin{tabularx}{\linewidth}{X}
    \toprule
    \textbf{PTKB classification.} \\
    \midrule
    I will give you some background information about a user and a conversation between the user and a system. You should tell me which of the background information is relevant for answering the last question of the user. \\
    Here is the background information about the user: \{$ptkb$\} \\
    Please just copy the relevant background information to the last user utterance. \\
    \bottomrule
\end{tabularx}
\label{tbl:prompt-ptkb}
\end{table}

%% file: main.bbl

\begin{thebibliography}{21}


\ifx \showCODEN    \undefined \def \showCODEN     #1{\unskip}     \fi
\ifx \showDOI      \undefined \def \showDOI       #1{#1}\fi
\ifx \showISBNx    \undefined \def \showISBNx     #1{\unskip}     \fi
\ifx \showISBNxiii \undefined \def \showISBNxiii  #1{\unskip}     \fi
\ifx \showISSN     \undefined \def \showISSN      #1{\unskip}     \fi
\ifx \showLCCN     \undefined \def \showLCCN      #1{\unskip}     \fi
\ifx \shownote     \undefined \def \shownote      #1{#1}          \fi
\ifx \showarticletitle \undefined \def \showarticletitle #1{#1}   \fi
\ifx \showURL      \undefined \def \showURL       {\relax}        \fi
\providecommand\bibfield[2]{#2}
\providecommand\bibinfo[2]{#2}
\providecommand\natexlab[1]{#1}
\providecommand\showeprint[2][]{arXiv:#2}

\bibitem[Abbasiantaeb et~al\mbox{.}(2024)]%
        {mq4cs}
\bibfield{author}{\bibinfo{person}{Zahra Abbasiantaeb}, \bibinfo{person}{Simon Lupart}, {and} \bibinfo{person}{Mohammad Aliannejadi}.} \bibinfo{year}{2024}\natexlab{}.
\newblock \showarticletitle{Generating Multi-Aspect Queries for Conversational Search}.
\newblock \bibinfo{journal}{\emph{arXiv preprint arXiv:2403.19302}} (\bibinfo{year}{2024}).
\newblock


\bibitem[Abbasiantaeb et~al\mbox{.}(2023)]%
        {abbasiantaeb2023llm}
\bibfield{author}{\bibinfo{person}{Zahra Abbasiantaeb}, \bibinfo{person}{Chuan Meng}, \bibinfo{person}{David Rau}, \bibinfo{person}{Antonis Krasakis}, \bibinfo{person}{Hossein~A Rahmani}, {and} \bibinfo{person}{Mohammad Aliannejadi}.} \bibinfo{year}{2023}\natexlab{}.
\newblock \showarticletitle{LLM-based Retrieval and Generation Pipelines for TREC Interactive Knowledge Assistance Track (iKAT) 2023}.
\newblock  (\bibinfo{year}{2023}).
\newblock


\bibitem[Aliannejadi et~al\mbox{.}(2024)]%
        {ikat23}
\bibfield{author}{\bibinfo{person}{Mohammad Aliannejadi}, \bibinfo{person}{Zahra Abbasiantaeb}, \bibinfo{person}{Shubham Chatterjee}, \bibinfo{person}{Jeffrey Dalton}, {and} \bibinfo{person}{Leif Azzopardi}.} \bibinfo{year}{2024}\natexlab{}.
\newblock \showarticletitle{TREC iKAT 2023: A Test Collection for Evaluating Conversational and Interactive Knowledge Assistants}. In \bibinfo{booktitle}{\emph{Proceedings of the 47th International ACM SIGIR Conference on Research and Development in Information Retrieval}} (Washington DC, USA) \emph{(\bibinfo{series}{SIGIR '24})}. \bibinfo{publisher}{Association for Computing Machinery}, \bibinfo{address}{New York, NY, USA}, \bibinfo{pages}{819–829}.
\newblock
\showISBNx{9798400704314}
\urldef\tempurl%
\url{https://doi.org/10.1145/3626772.3657860}
\showDOI{\tempurl}


\bibitem[Clark(2020)]%
        {clark2020electra}
\bibfield{author}{\bibinfo{person}{K Clark}.} \bibinfo{year}{2020}\natexlab{}.
\newblock \showarticletitle{Electra: Pre-training text encoders as discriminators rather than generators}.
\newblock \bibinfo{journal}{\emph{arXiv preprint arXiv:2003.10555}} (\bibinfo{year}{2020}).
\newblock


\bibitem[Dalton et~al\mbox{.}(2020a)]%
        {cast20}
\bibfield{author}{\bibinfo{person}{Jeffrey Dalton}, \bibinfo{person}{Chenyan Xiong}, {and} \bibinfo{person}{Jamie Callan}.} \bibinfo{year}{2020}\natexlab{a}.
\newblock \showarticletitle{CAsT 2020: The Conversational Assistance Track Overview}. In \bibinfo{booktitle}{\emph{Text Retrieval Conference}}.
\newblock
\urldef\tempurl%
\url{https://api.semanticscholar.org/CorpusID:214735659}
\showURL{%
\tempurl}


\bibitem[Dalton et~al\mbox{.}(2020b)]%
        {cast19}
\bibfield{author}{\bibinfo{person}{Jeffrey Dalton}, \bibinfo{person}{Chenyan Xiong}, \bibinfo{person}{Vaibhav Kumar}, {and} \bibinfo{person}{Jamie Callan}.} \bibinfo{year}{2020}\natexlab{b}.
\newblock \showarticletitle{CAsT-19: A Dataset for Conversational Information Seeking}. In \bibinfo{booktitle}{\emph{Proceedings of the 43rd International ACM SIGIR Conference on Research and Development in Information Retrieval}} (Virtual Event, China) \emph{(\bibinfo{series}{SIGIR '20})}. \bibinfo{publisher}{Association for Computing Machinery}, \bibinfo{address}{New York, NY, USA}, \bibinfo{pages}{1985–1988}.
\newblock
\showISBNx{9781450380164}
\urldef\tempurl%
\url{https://doi.org/10.1145/3397271.3401206}
\showDOI{\tempurl}


\bibitem[Formal et~al\mbox{.}(2021a)]%
        {formal2021splade}
\bibfield{author}{\bibinfo{person}{Thibault Formal}, \bibinfo{person}{Carlos Lassance}, \bibinfo{person}{Benjamin Piwowarski}, {and} \bibinfo{person}{St{\'e}phane Clinchant}.} \bibinfo{year}{2021}\natexlab{a}.
\newblock \showarticletitle{SPLADE v2: Sparse lexical and expansion model for information retrieval}.
\newblock \bibinfo{journal}{\emph{arXiv preprint arXiv:2109.10086}} (\bibinfo{year}{2021}).
\newblock


\bibitem[Formal et~al\mbox{.}(2022)]%
        {10.1145/3477495.3531857/splade++}
\bibfield{author}{\bibinfo{person}{Thibault Formal}, \bibinfo{person}{Carlos Lassance}, \bibinfo{person}{Benjamin Piwowarski}, {and} \bibinfo{person}{St\'{e}phane Clinchant}.} \bibinfo{year}{2022}\natexlab{}.
\newblock \showarticletitle{From Distillation to Hard Negative Sampling: Making Sparse Neural IR Models More Effective}. In \bibinfo{booktitle}{\emph{Proceedings of the 45th International ACM SIGIR Conference on Research and Development in Information Retrieval}} (Madrid, Spain) \emph{(\bibinfo{series}{SIGIR '22})}. \bibinfo{publisher}{Association for Computing Machinery}, \bibinfo{address}{New York, NY, USA}, \bibinfo{pages}{2353–2359}.
\newblock
\showISBNx{9781450387323}
\urldef\tempurl%
\url{https://doi.org/10.1145/3477495.3531857}
\showDOI{\tempurl}


\bibitem[Formal et~al\mbox{.}(2021b)]%
        {spladev1}
\bibfield{author}{\bibinfo{person}{Thibault Formal}, \bibinfo{person}{Benjamin Piwowarski}, {and} \bibinfo{person}{St\'{e}phane Clinchant}.} \bibinfo{year}{2021}\natexlab{b}.
\newblock \showarticletitle{SPLADE: Sparse Lexical and Expansion Model for First Stage Ranking}. In \bibinfo{booktitle}{\emph{Proceedings of the 44th International ACM SIGIR Conference on Research and Development in Information Retrieval}} (Virtual Event, Canada) \emph{(\bibinfo{series}{SIGIR '21})}. \bibinfo{publisher}{Association for Computing Machinery}, \bibinfo{address}{New York, NY, USA}, \bibinfo{pages}{2288–2292}.
\newblock
\showISBNx{9781450380379}
\urldef\tempurl%
\url{https://doi.org/10.1145/3404835.3463098}
\showDOI{\tempurl}


\bibitem[He et~al\mbox{.}(2021)]%
        {he2021debertav3}
\bibfield{author}{\bibinfo{person}{Pengcheng He}, \bibinfo{person}{Jianfeng Gao}, {and} \bibinfo{person}{Weizhu Chen}.} \bibinfo{year}{2021}\natexlab{}.
\newblock \showarticletitle{Debertav3: Improving deberta using electra-style pre-training with gradient-disentangled embedding sharing}.
\newblock \bibinfo{journal}{\emph{arXiv preprint arXiv:2111.09543}} (\bibinfo{year}{2021}).
\newblock


\bibitem[He et~al\mbox{.}(2020)]%
        {he2020deberta}
\bibfield{author}{\bibinfo{person}{Pengcheng He}, \bibinfo{person}{Xiaodong Liu}, \bibinfo{person}{Jianfeng Gao}, {and} \bibinfo{person}{Weizhu Chen}.} \bibinfo{year}{2020}\natexlab{}.
\newblock \showarticletitle{Deberta: Decoding-enhanced bert with disentangled attention}.
\newblock \bibinfo{journal}{\emph{arXiv preprint arXiv:2006.03654}} (\bibinfo{year}{2020}).
\newblock


\bibitem[J\"{a}rvelin and Kek\"{a}l\"{a}inen(2000)]%
        {ndcg}
\bibfield{author}{\bibinfo{person}{Kalervo J\"{a}rvelin} {and} \bibinfo{person}{Jaana Kek\"{a}l\"{a}inen}.} \bibinfo{year}{2000}\natexlab{}.
\newblock \showarticletitle{IR Evaluation Methods for Retrieving Highly Relevant Documents}. In \bibinfo{booktitle}{\emph{Proceedings of the 23rd Annual International ACM SIGIR Conference on Research and Development in Information Retrieval}} (Athens, Greece) \emph{(\bibinfo{series}{SIGIR '00})}. \bibinfo{publisher}{ACM}, \bibinfo{address}{New York, NY, USA}, \bibinfo{pages}{41--48}.
\newblock
\showISBNx{1-58113-226-3}
\urldef\tempurl%
\url{https://doi.org/10.1145/345508.345545}
\showDOI{\tempurl}


\bibitem[Lan(2019)]%
        {lan2019albert}
\bibfield{author}{\bibinfo{person}{Z Lan}.} \bibinfo{year}{2019}\natexlab{}.
\newblock \showarticletitle{Albert: A lite bert for self-supervised learning of language representations}.
\newblock \bibinfo{journal}{\emph{arXiv preprint arXiv:1909.11942}} (\bibinfo{year}{2019}).
\newblock


\bibitem[Lassance and Clinchant(2023)]%
        {lassance2023naver}
\bibfield{author}{\bibinfo{person}{Carlos Lassance} {and} \bibinfo{person}{St{\'e}phane Clinchant}.} \bibinfo{year}{2023}\natexlab{}.
\newblock \showarticletitle{Naver Labs Europe (SPLADE)@ TREC Deep Learning 2022}.
\newblock \bibinfo{journal}{\emph{arXiv preprint arXiv:2302.12574}} (\bibinfo{year}{2023}).
\newblock


\bibitem[Lassance et~al\mbox{.}(2024)]%
        {lassance2024spladev3}
\bibfield{author}{\bibinfo{person}{Carlos Lassance}, \bibinfo{person}{Hervé Déjean}, \bibinfo{person}{Thibault Formal}, {and} \bibinfo{person}{Stéphane Clinchant}.} \bibinfo{year}{2024}\natexlab{}.
\newblock \bibinfo{title}{SPLADE-v3: New baselines for SPLADE}.
\newblock
\newblock


\bibitem[Lewis et~al\mbox{.}(2020)]%
        {RAG-first-work}
\bibfield{author}{\bibinfo{person}{Patrick Lewis}, \bibinfo{person}{Ethan Perez}, \bibinfo{person}{Aleksandra Piktus}, \bibinfo{person}{Fabio Petroni}, \bibinfo{person}{Vladimir Karpukhin}, \bibinfo{person}{Naman Goyal}, \bibinfo{person}{Heinrich K\"{u}ttler}, \bibinfo{person}{Mike Lewis}, \bibinfo{person}{Wen-tau Yih}, \bibinfo{person}{Tim Rockt\"{a}schel}, \bibinfo{person}{Sebastian Riedel}, {and} \bibinfo{person}{Douwe Kiela}.} \bibinfo{year}{2020}\natexlab{}.
\newblock \showarticletitle{Retrieval-augmented generation for knowledge-intensive NLP tasks}. In \bibinfo{booktitle}{\emph{Proceedings of the 34th International Conference on Neural Information Processing Systems}} (Vancouver, BC, Canada) \emph{(\bibinfo{series}{NIPS '20})}. \bibinfo{publisher}{Curran Associates Inc.}, \bibinfo{address}{Red Hook, NY, USA}, Article \bibinfo{articleno}{793}, \bibinfo{numpages}{16}~pages.
\newblock
\showISBNx{9781713829546}


\bibitem[Liu(2019)]%
        {liu2019roberta}
\bibfield{author}{\bibinfo{person}{Yinhan Liu}.} \bibinfo{year}{2019}\natexlab{}.
\newblock \showarticletitle{Roberta: A robustly optimized bert pretraining approach}.
\newblock \bibinfo{journal}{\emph{arXiv preprint arXiv:1907.11692}}  \bibinfo{volume}{364} (\bibinfo{year}{2019}).
\newblock


\bibitem[Lupart et~al\mbox{.}(2024)]%
        {lupart2024disco}
\bibfield{author}{\bibinfo{person}{Simon Lupart}, \bibinfo{person}{Mohammad Aliannejadi}, {and} \bibinfo{person}{Evangelos Kanoulas}.} \bibinfo{year}{2024}\natexlab{}.
\newblock \showarticletitle{DiSCo Meets LLMs: A Unified Approach for Sparse Retrieval and Contextual Distillation in Conversational Search}.
\newblock \bibinfo{journal}{\emph{arXiv preprint arXiv:2410.14609}} (\bibinfo{year}{2024}).
\newblock


\bibitem[Mao et~al\mbox{.}(2023)]%
        {mao2023largelanguagemodelsknowllm4cs}
\bibfield{author}{\bibinfo{person}{Kelong Mao}, \bibinfo{person}{Zhicheng Dou}, \bibinfo{person}{Fengran Mo}, \bibinfo{person}{Jiewen Hou}, \bibinfo{person}{Haonan Chen}, {and} \bibinfo{person}{Hongjin Qian}.} \bibinfo{year}{2023}\natexlab{}.
\newblock \showarticletitle{Large Language Models Know Your Contextual Search Intent: A Prompting Framework for Conversational Search}. In \bibinfo{booktitle}{\emph{Findings of the Association for Computational Linguistics: EMNLP 2023}}, \bibfield{editor}{\bibinfo{person}{Houda Bouamor}, \bibinfo{person}{Juan Pino}, {and} \bibinfo{person}{Kalika Bali}} (Eds.). \bibinfo{publisher}{Association for Computational Linguistics}, \bibinfo{address}{Singapore}, \bibinfo{pages}{1211--1225}.
\newblock
\urldef\tempurl%
\url{https://doi.org/10.18653/v1/2023.findings-emnlp.86}
\showDOI{\tempurl}


\bibitem[Mo et~al\mbox{.}(2024)]%
        {mo2024chiqcontextualhistoryenhancement}
\bibfield{author}{\bibinfo{person}{Fengran Mo}, \bibinfo{person}{Abbas Ghaddar}, \bibinfo{person}{Kelong Mao}, \bibinfo{person}{Mehdi Rezagholizadeh}, \bibinfo{person}{Boxing Chen}, \bibinfo{person}{Qun Liu}, {and} \bibinfo{person}{Jian-Yun Nie}.} \bibinfo{year}{2024}\natexlab{}.
\newblock \showarticletitle{CHIQ: Contextual History Enhancement for Improving Query Rewriting in Conversational Search}.
\newblock \bibinfo{journal}{\emph{arXiv preprint arXiv:2406.05013}} (\bibinfo{year}{2024}).
\newblock


\bibitem[Owoicho et~al\mbox{.}(2022)]%
        {cast22}
\bibfield{author}{\bibinfo{person}{Paul Owoicho}, \bibinfo{person}{Jeffrey Dalton}, \bibinfo{person}{Mohammad Aliannejadi}, \bibinfo{person}{Leif Azzopardi}, \bibinfo{person}{Johanne~R. Trippas}, {and} \bibinfo{person}{Svitlana Vakulenko}.} \bibinfo{year}{2022}\natexlab{}.
\newblock \showarticletitle{TREC CAsT 2022: Going Beyond User Ask and System Retrieve with Initiative and Response Generation}. In \bibinfo{booktitle}{\emph{Text Retrieval Conference}}.
\newblock
\urldef\tempurl%
\url{https://api.semanticscholar.org/CorpusID:261288646}
\showURL{%
\tempurl}


\end{thebibliography}
